\documentclass[universe,article,accept,moreauthors,pdftex,10pt,a4paper]{Definitions/mdpi} 
\pdfoutput=1
\firstpage{1} 
\pubvolume{xx}
\issuenum{1}
\articlenumber{5}
\pubyear{2019}
\copyrightyear{2019}
\history{Received: 28 November 2018; Accepted: 15 January 2019; Published: date}
\updates{yes} 

\Title{Entanglement and Disordered-Enhanced Topological Phase in the Kitaev Chain$^{\dagger}$}

\Author{Liron Levy and Moshe Goldstein *}

\AuthorNames{Liron Levy and Moshe Goldstein}

\address[1]{%
Raymond and Beverly Sackler School of Physics and Astronomy, Tel Aviv University, Tel Aviv 6997801, Israel; r3dr4gon@gmail.com}

 \corres{\hangafter=1 \hangindent=1.05em \hspace{-0.82em}Correspondence: mgoldstein@tauex.tau.ac.il}
\firstnote{This paper is based on a talk at the 7th International Conference on New Frontiers in Physics (ICNFP 2018), Crete, Greece, 4--12 July 2018.}

\abstract{In recent years, tools from quantum information theory have become indispensable in characterizing many-body systems. In this work, we employ measures of entanglement to study the interplay between disorder and the topological phase in 1D systems of the Kitaev type, which can host Majorana end modes at their edges. We find that the entanglement entropy may actually increase as a result of disorder, and identify the origin of this behavior in the appearance of an infinite-disorder critical point. We also employ the entanglement spectrum to accurately determine the phase diagram of the system, and find that disorder may enhance the topological phase, and lead to the appearance of Majorana zero modes in systems whose clean version is trivial.}


\begin{document}


\section{Introduction} \label{sec:introduction}
Entanglement is the quintessential characteristic of the quantum world. Recently, much attention has been devoted to manifestations of entanglement in quantum many-body systems~\cite{amico08,laflorencie2016}. Motivations include the prospect of using many-body ground and excited states as a resource for quantum communication and computation; the amount of entanglement controls the applicability of a powerful tensor-network-based numerical method; entanglement is an indicator of quantum correlations which is independent of the details of the system, allowing, in particular, one to identify quantum phase transitions even without knowing the relevant order parameter. In this work, we will concentrate on the latter property, in the context of disordered 1D quantum many-body systems.

Majorana zero modes~\cite{kitaev01,alicea12,sato16,aguado17} and their more complicated relatives~\cite{alicea16} have been studied extensively recently, both due to their predicted exotic non-abelian braiding statistics, which gives rise to the prospect of topologically-protected quantum computation~\cite{sarma15}, and to many concrete proposals for their realization in the laboratory. 
Leading candidates are semiconductor quantum wires with strong spin orbit interaction, which are rendered effectively spinless by the application of an appropriate magnetic field and gate voltage, and driven into a topological phase by proximity-coupling to a superconductor, leading to the formation of Majorana end modes~\cite{lutchyn10,oreg10}. Indications for these modes have recently been measured experimentally~\cite{mourik12,deng12,das12,finck13,churchill13}. These Majorana zero modes show up not only in the presence of a real edge, but also in the entanglement spectrum, that is, in the spectrum of the reduced density matrix of a subsystem~\cite{li08}.

Disorder naturally occurs in all these systems, and may hamper their topological characteristics, especially the Majorana edge modes of 1D systems~\cite{motrunich01,gruzberg05,akhmerov11,fulga11,potter11,stanescu11,brouwer11a,brouwer11b,sau12,lobos12,pientka12,pientka13,degottardi13a,neven13,rieder13,sau13,degottardi13b,chevallier13,jacquod13,adagideli14,hui14,crepin14,gregs16,hedge16,bagrets16,grabsch16,pekerten17,brzezicki17,mcginley17,lieu18,monthus18,wang18,mishra18}.
In this work, we will study the interplay of disorder and the topological phase from the entanglement point of view. After introducing the model and our method for calculating the entanglement spectrum and entropy in Section~\ref{sec:model}, we will examine the entanglement entropy in Section~\ref{sec:entropy}. We will show that although superconducting proximity and disorder each tend to separately suppress the entanglement, their interplay may enhance the entanglement and cause it to behave in a non-monotonic fashion. We will explain this as the result of a strong-disorder quantum criticality.
In Section~\ref{sec:phase}, we will use the entanglement spectrum, and in particular, the presence of entanglement Majorana zero modes, to distinguish between the two phases (topological and trivial). We will show that in certain regimes of the parameter, space disorder can actually enhance or even be the origin of the topological phase in the system. We will summarize our findings and give a future outlook in Section~\ref{sec:conclusions}.

\section{Model and Method} \label{sec:model}
We will study the standard Kitaev chain~\cite{kitaev01} model with disorder. The model describes spinless fermions hopping on a 1D lattice in the presence of (proximity-induced) pairing and disorder potential,
\begin{equation} \label{eqn:h_kitaev}
	H_\mathrm{Kitaev} = \sum_{n=1}^L (V_n - \mu) c_n^\dagger c_n -t c_{n+1}^\dagger c_n - \Delta c_{n+1} c_n + \mathrm{h.c.},
\end{equation}
where $c_n^\dagger$ creates a fermion (with the standard anticommutation relations) at site $n=1 \ldots L$ (and $L+1 \equiv 1$ if the boundary conditions are periodic; for open boundary conditions, terms which refer to site $L+1$ are omitted), $V_n$ is the disorder potential, independently and uniformly distributed in the interval $[-W/2,W/2]$, $\mu$ is the chemical potential, and $t$ and $\Delta$ are, respectively, the hopping matrix element and the pairing amplitude, which one may choose as real without loss of generality.

Since the model is quadratic, it can be solved using a Bogolubov transformation.
First, let us define the Nambu--Gorkov Fermi operators, $\psi_{i}$ ($i = 1 \cdots 2L$) as $\psi_{n}=c_i$ and $\psi_{n+L} = c_n^\dagger$ for $n=1 \cdots L$. Then, the Hamiltonian has the general Bogolubov--de Gennes form
\begin{equation} \label{eqn:h_bogolubov}
	H_\mathrm{BdG} = \sum_{i,j=1}^{2L} h_{ij} \psi_i^\dagger \psi_j,
\end{equation}
where in our specific case the only nonzero elements of the $2L \times 2L$ matrix $h_{ij}$ are $h_{nn} = - h_{n+L,n+L}  = v_n - \mu$, $h_{n,n+1} = h_{n+1,n} = -h_{n+L,n+L+1} = - h_{n+L+1,n+L} = -t$ and $h_{n,n+L+1} = -h_{n+1,n+L} = -h_{n+L,n+1} = h_{n+L+1,n} = -\Delta$.
The problem then reduces to the diagonalization of the ``single-particle Hamiltonian'' matrix $h_{ij}$:
Let us denote by $U_{ij}$ the unitary matrix whose $j$th column contains the $j$th normalized eigenvector of the matrix $h_{ij}$, corresponding to the real eigenvalue $\lambda_j$ (which are guaranteed to come in pairs of equal magnitude and opposite sign), so that $\sum_{k=1}^{2L} h_{ik} U_{kj} = U_{kj} \lambda_j$ ($i,j,k=1 \cdots 2L$). Defining $\tilde{\psi}_i = \sum_{j=1}^{2L} U_{ji}^* \psi_j$ (so that $\psi_i = \sum_{j=1}^{2L} U_{ij} \tilde{\psi}_j$), one has $H = \sum_{i=1}^{2L} \lambda_i \tilde{\psi}^\dagger_i \tilde{\psi}_i$.

In the clean case ($V_n=0$), the model is easily solved analytically~\cite{kitaev01}. It has two phases, depending on the ratio $|\mu/t|$: For $|\mu/t|>2$, it is a topologically trivial insulator (qualitatively similar to the limit $|\mu/t| \to \infty$, where the system is either completely filled or completely empty, depending on the sign of $\mu$). For $|\mu/t|<2$, the system is a topological superconductor, which features a single Majorana zero mode exponentially localized at each physical boundary (hence their hybridization is exponentially small in the system size), with the localization length inversely proportional to the energy gap. At $|\mu/t|=2$, the gap closes and the system features a 1D Majorana mode which allows the boundary zero modes to merge and disappear.

Let us now return to the general case, and discuss the calculation of observables in the many-body ground state of the model. In this state
(to be denoted by $|\Phi_0\rangle$), all the modes with $\lambda_i<0$ are occupied. Therefore, $\langle \tilde{\psi}_i^\dagger  \tilde{\psi}_j \rangle = \delta_{ij} \theta (-\lambda_j)$, where $\theta$ is Heaviside's step function, and hence $c_{ij} = \langle \psi_i^\dagger \psi_j \rangle = \sum_k U_{ik}^* U_{kj} \theta(-\lambda_k)$.
Our main interest is in the entanglement between a spatial subsystem $A$ (defined by some subset of sites $A \subset \{1,\cdots,L\}$ of size $|A|$) and its complement $\bar{A}$, when the total system is in its ground state $|\Phi_0\rangle$.
In this case, the entanglement can be determined from the mixedness of its reduced many-body density matrix,
$\rho_A = \mathrm{Tr}_{\bar{A}} |\Phi_0\rangle \langle\Phi_0|$.
Since the system is quadratic, $\rho_A$ is Gaussian, i.e., it can be written as $\rho_A \propto \exp(-H_A)$ with a quadratic ``entanglement Hamiltonian''~\cite{peschel03}
\begin{equation} \label{eqn:h_entanglement}
H_A=\sum_{i,j \in A \cup A+L} h_{A,ij} \psi_i^\dagger \psi_j,
\end{equation}
where the dimensions of the matrix $h_{A,ij}$ are $2|A| \times 2|A|$. 
By Wick's theorem, the latter is completely characterized by the single-time two-point fermionic correlation function, $c_{A,ij} = \langle \psi_i^\dagger \psi_j \rangle$, which is simply the submatrix of $c_{ij}$ of size $2|A| \times 2|A|$, with $i,j \in A \cup A + L$~\cite{peschel03}. The latter is related to the matrix $h_A$ by $h_A = \ln [(1-c_A)/c_A]$. This allows one to extract the entire spectrum of $h_A$ or $\rho_A$, as well as its moments. In particular, the von~Neumann entanglement entropy, $S_A = -\mathrm{Tr} (\rho_A \ln \rho_A) = S_{\bar{A}}$, can be expressed in terms of the $2|A|$ eigenvalues $f_{A,i}$ of the matrix $c_A$ as~\cite{peschel03}
\begin{equation}
S_A = - \sum_i f_{A,i} \ln f_{A,i}. 
\end{equation}

\section{Entanglement Entropy} \label{sec:entropy}
We will start by studying the von Neumann entanglement entropy of a subsystem $A$ which consists of the sites $n=1,\cdots,L_A$ (a single continuous interval of length $L_A<L/2$).
Generally in 1D, if the system is critical (gapless) and not localized by disorder, one expects the entanglement entropy to scale logarithmically with the subsystem size~\cite{holzhey94,vidal03,calabrese04,calabrese09},
\begin{equation} \label{eqn:s_cft}
  S_A \sim \frac{c}{3} \ln(L_A) + \mathrm{const.},
\end{equation}
where $c$ is the central charge of the corresponding conformal field theory, which roughly counts the number of gapless 1D modes in the system. For example, $c=1/2$ for a non-chiral Majorana (real) fermion mode and $c=1$ for a non-chiral Dirac (complex) fermion, or, equivalently, a non-chiral gapless bosonic mode. The above expression is valid when the interval in question has two boundaries with the rest of the system; for open boundary conditions with subsystem A located at one end, the coefficient of the logarithm is halved.
The same logarithmic scaling holds as a function of the total system size $L$ if the ratio $L_A/L$ is kept fixed.
In the presence of a gap, this logarithmic scaling holds up to the correlation length (inversely proportional to the gap), above which the entanglement entropy saturates. Similar behavior holds for a gapless disordered localized system, with the role of correlation length being taken by the localization length.

\begin{figure}[H]
	\centering
	\includegraphics[width=\textwidth,clip]{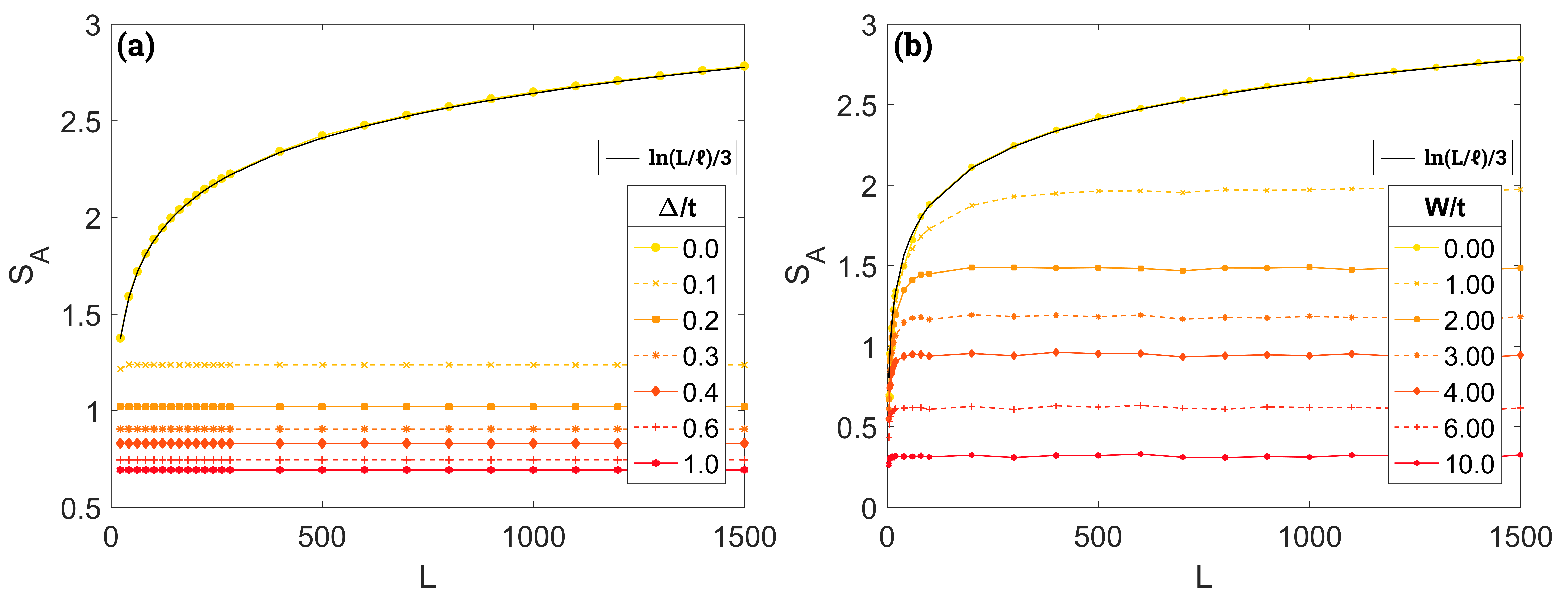}
	\caption{von Neumann entanglement entropy (disorder-averaged over $10^4$ samples when $W > 0$) as a function of the total system size $L$ for $L_A/L=1/2$, $\mu=0$, and periodic boundary conditions: (\textbf{a}) In the clean case ($W=0$) for different values of $\Delta/t$; (\textbf{b}) For $\Delta=0$ and different disorder strengths $W/t$. For $W=\Delta=0$, a fit to the expected logarithmic critical behavior, Equation~(\ref{eqn:s_cft}), is plotted in black.}
	\label{fig:entropy_separate}
\end{figure}

\begin{figure}[t]
	\centering
	\includegraphics[width=\textwidth,clip]{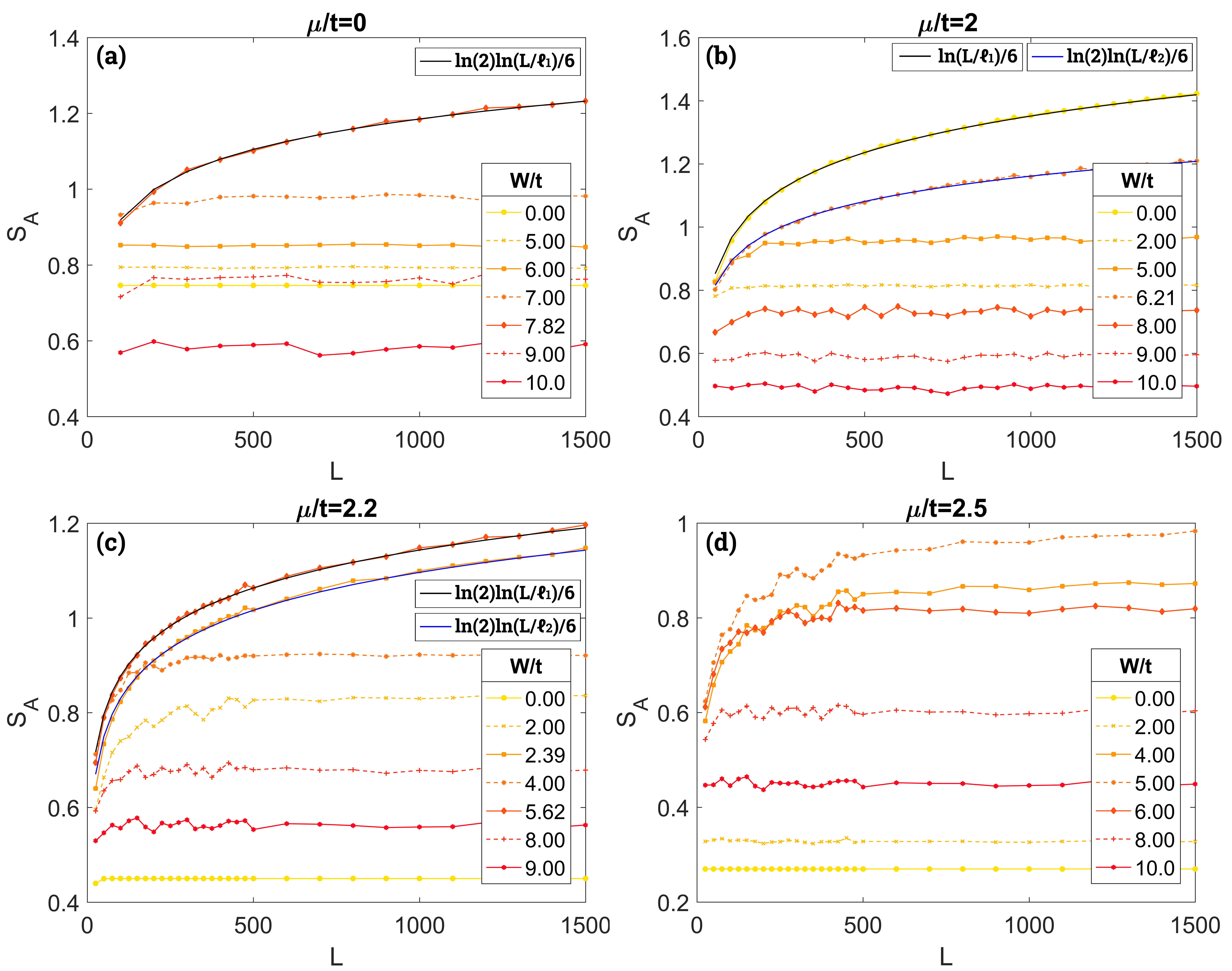}
	\caption{von Neumann entanglement entropy (disorder-averaged over $10^4$ samples when $W>0$) as a function of the total system size $L$ for $L_A/L=1/2$, $\Delta/t=0.6$, and periodic boundary conditions, for different values of the disorder strength $W/t$. Different panels correspond to different chemical potential: (\textbf{a}) $\mu/t=0$; (\textbf{b}) $\mu/t=2$; (\textbf{c}) $\mu/t=2.2$; (\textbf{d}) $\mu/t=2.5$. Black and blue curves indicate fits to a logarithmic critical behavior, as in Equation~(\ref{eqn:s_cft}).}
	\label{fig:entropy_mu}
\end{figure}

Let us examine the behavior of the entanglement entropy in our system as a function of the total system size $L$. We will henceforth consider only periodic boundary conditions; open boundaries were found to give similar effects (up to the above-mentioned factor of 2).
In Figure~\ref{fig:entropy_separate}, we concentrate on $\mu=0$ and start from the clean case ($W=0$) without superconductivity ($\Delta=0$). Then, we have just a half-filled band of free fermions, and one finds the expected logarithmic scaling, $S_A \sim (c/3)\ln(L)$ with $c=1$ (gapless Dirac fermion). Adding either superconductivity ($\Delta \ne 0$) or disorder ($W>0$) separately makes the entanglement saturate as a function of $L$, due to either the finite correlation length (inversely proportional to the gap $\Delta$)
or localization length (which decreases as $W$ increases).
One would then expect that if one introduces both superconductivity and disorder together, the entanglement entropy will saturate even more quickly. Interestingly, this is not the case: as can be seen in Figure~\ref{fig:entropy_mu}{a}, if we fix $\Delta$ to a finite value and increase $W$ from zero, the entanglement starts to increase at some value of $W$, reaching a logarithmic behavior at $W/t \approx 7.82$ (the way to accurately determine this value will be discussed in the next section), before decreasing again as $W$ is further increased.

To understand the origin of this peculiar behavior, it is better to employ a dual point of view. Let us concentrate on the case $\Delta=t$. Using a Jordan--Wigner transformation to define standard spin-1/2 operators at each site $n$ (presented for open boundary conditions, but could be adapted to periodic ones as well),
\begin{align} \label{eqn:ising}
S^x_n & = c_n^\dagger c_n - 1/2, \\
S^z_n & = (c_n^\dagger + c_n) \prod_{n^\prime=1}^{n-1} (2c_{n^\prime}^\dagger c_{n^\prime} - 1) ,
\end{align}
the Hamiltonian~(\ref{eqn:h_kitaev}) can be mapped into that of the disordered transverse-field 1D Ising model,
\begin{equation}
  H_\mathrm{Ising} = \sum_{n=1}^L h_n S^x_n - J_n S^z_n S^z_{n+1},
\end{equation}
where $h_n=(V_n-\mu)/2$, $J_n = t = \Delta$. This model has a $\mathbb{Z}_2$ symmetry $S^z_n \to -S^z_n$, corresponding to the overall fermion parity in the Kitaev chain.
In the absence of disorder ($W=0$), this Hamiltonian has two phases: when $|h|>|J|$, the Zeeman term dominates, resulting in a paramagnetic phase; when $|h|<|J|$, the exchange term dominates, resulting in a broken-symmetry ferromagnetic phase (for $J>0$). The non-local Jordan--Wigner transformation maps these into the (symmetry-unbroken) trivial and topological superconductor phase of the Kitaev chain, respectively.
However, now we can understand that the same phases exist even when $J_n$ and/or $h_n$ become random: when the former dominates, a ferromagnet still results (if the sign of $J_n$ is fixed, it is a usual ferromagnet; if not, one may flip $S^z_{n^\prime}$ between each pair of negative $J_n$ to make all the $J_n$ positive, so the original system is a ferromagnet in disguise), while when the latter dominates a paramagnet appears~\cite{ma79,dasgupta80,fisher94,fisher95}.

More interesting is the transition between the two phases in the disordered case, which presents a new type of quantum criticality, known as the infinite-disorder critical point~\cite{ma79,dasgupta80,fisher94,fisher95}. Its physics has been analyzed using the strong disorder renormalization group. In the latter, at each step, one picks the largest coupling in the chain. If this is a magnetic field $h_n$, the corresponding spin is polarized according to the sign of $h_n$; its neighboring spins are coupled by a new exchange term, $\sim J_{n-1} J_{n}/h_n$. If, on the other hand, the largest coupling is an exchange term $J_n$, the spins coupled by it become restricted to the two-dimensional low energy space $|\Uparrow\rangle \equiv |\uparrow\uparrow\rangle$ and $|\Downarrow\rangle \equiv |\downarrow\downarrow\rangle$ (assuming $J_n>0$), which can be thought of as a new effective spin-1/2. This effective spin-1/2 experiences an effective magnetic field $\sim h_n h_{n+1}/J_n$. If, at a later stage of the renormalization, this spin becomes polarized in the $x$ direction, a Bell state is formed in terms of the original spins. Counting how many such Bell states straddle the boundary between subsystem A and the rest of the system, one finds that the entanglement entropy features a logarithmic behavior, as in Equation~(\ref{eqn:s_cft}), but with an irrational coefficient, $c_\mathrm{eff}=\ln 2/2$~\cite{refael04,refael09}. This is exactly the behavior that we observed in Figure~\ref{fig:entropy_mu}{a} at $W/t\approx 7.82$.

This also has a natural interpretation from the original fermion point of view~\cite{motrunich01,gruzberg05,brouwer11b}. In the clean topological phase, Majorana zero modes exist at the ends of the system. Due to the gap, their coupling and hence splitting are exponentially-small in the distance between them, that is, the system size. When disorder is introduced, bulk modes start filling in the gap. At some critical value of the disorder, these gap modes approach zero energy, creating a channel which allows the Majoranas at the ends to couple and eliminate each other, making the system trivial. This transition point corresponds exactly to the strong-disorder critical point in the spin chain language. The fermionic picture also shows that the transition should persist when $\Delta \ne t$, where the equivalent spin chain becomes more complicated than the disordered transverse-field Ising model, Equation~(\ref{eqn:ising}). This is supported by our results in Figure~\ref{fig:entropy_mu}, where $\Delta/t=0.6 \ne 1$.

Let us continue examining our data. New intriguing behavior arises when one introduces a finite chemical potential $\mu$. Figure~\ref{fig:entropy_mu}{b} depicts the behavior for $\mu/t=2$. Here, in the clean limit, $W=0$, the system is at the transition from a topological superconductor to a trivial state, featuring a logarithmic behavior described by Equation~(\ref{eqn:s_cft}) with $c=1/2$, due to the Majorana nature of the critical (gapless) bulk mode. One would then na\"{\i}vly expect the disorder (which suppressed the topological phase at $\mu=0$) to immediately drive the system into the trivial phase. However, in reality, while the entanglement entropy initially drops with disorder, it then increases again to a logarithmic behavior with a $c_\mathrm{eff}=\ln 2/2$ coefficient at $W/t \approx 6.21$, before dropping back again. This shows that, in this case, disorder initially drives the system into the topological phase, and only later on makes it trivial, via a strong-disorder critical point. This behavior persists at $\mu/t=2.2$ (Figure~\ref{fig:entropy_mu}{c}), where the clean system is trivial, and increasing disorder causes the entanglement to grow to a logarithmic behavior at $W/t \approx 2.39$, then drop, increase again to a logarithmic behavior at $W/t \approx 5.62$, then drop again. In~both cases, $c_\mathrm{eff}=\ln 2/2$. Thus, here, disorder causes two transitions: from trivial to topological and back. Finally, at $\mu/t = 2.5$, logarithmic behavior is never obtained, indicating that the system stays trivial for all values of $W$ (Figure~\ref{fig:entropy_mu}{d}). All of this prompts one to study more carefully the phase diagram of the system, which will be the topic of the next section.

\begin{figure}[t]
	\centering
	\includegraphics[width=\textwidth,clip]{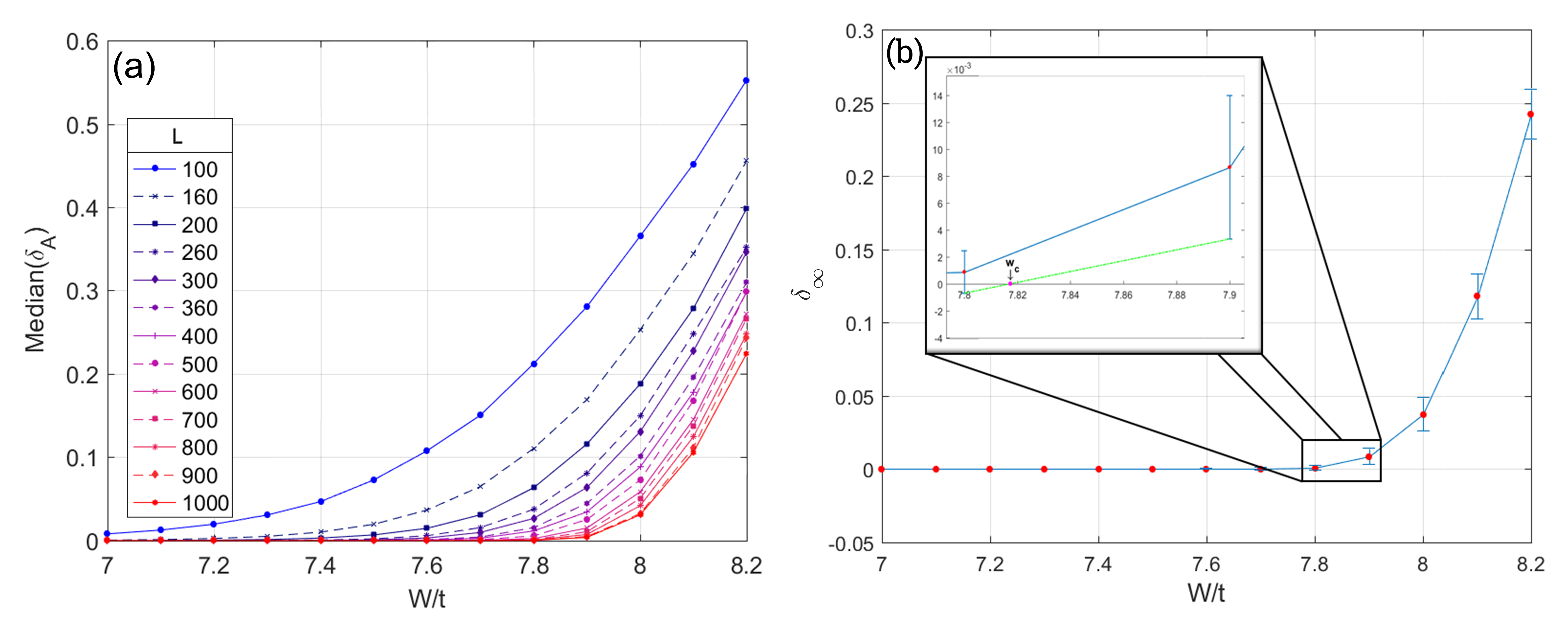}
	\caption{(\textbf{a}) The median (taken over $10^4$ realizations) of the gap $\delta_A$ between the first two eigenvalues of the single-particle entanglement spectrum [spectrum of the entanglement Hamiltonian, Equation~(\ref{eqn:h_entanglement})], as a function of the disorder $W/t$ for $L_A/L=1/2$, $\Delta/t=0.6$, $\mu=0$, and periodic boundary conditions, for different values of the total system size $L$. (\textbf{b}) The constant coefficient $\delta_{\infty}$ in fitting the $L$ dependence of $\delta_A$ to Equation~(\ref{eqn:deltaA_L}) as a function of the disorder strength $W/t$. The inset shows a blow-up of the region where $\delta_{\infty}$ switches from zero to nonzero. Our method of determining the critical disorder strength $W_c$ is also indicated---see the text for further details.}
	\label{fig:spectrum}       
\end{figure}

\section{Phase Diagram from Entanglement} \label{sec:phase}
The non-monotonic changes in the subsystem size dependence of the entanglement entropy as disorder is varied, discussed in the previous section, call for a more careful study of the phase diagram of the system. However, the $S_A(L)$ curves are not a very accurate means to probe the phase boundaries, since for reasonable system sizes it is not easy to distinguish numerically between a critical logarithmic behavior, Equation~(\ref{eqn:s_cft}), and a saturation at some constant value.

To overcome this difficulty, we employ a new strategy, looking at the entanglement spectrum, that is, the spectrum of the single-particle entanglement Hamiltonian, Equation~(\ref{eqn:h_entanglement}). In the topological phase, it should feature Majorana zero modes at the ends of the subsystem, just like a system with open ends, whereas in the trivial phase there should be no such zero modes. This is exemplified in Figure~\ref{fig:spectrum}{a}, where the median of the gap $\delta_A$ between the two lowest eigenvalues of the entanglement Hamiltonian~(\ref{eqn:h_entanglement}) is plotted as a function of disorder for different system sizes. One can clearly distinguish between fast decay to zero for small $W/t$ and apparent saturation for large $W/t$, as the subsystem size $L_A$ is increased. To make this quantitative, for each disorder strength we fit the dependence of $\delta_A$ on the subsystem size to:
\begin{equation} \label{eqn:deltaA_L}
  \delta_A (L_A) = \delta_{\infty} + \delta_{0} e^{-L_A/\xi_\delta}.
\end{equation}

We would like to identify whether the constant term $\delta_\infty$ is finite or zero.
For this purpose, it is very important to use the correct form for the second, length-dependent, term in the case where the constant $\delta_{\infty}$ is zero, but not when $\delta_{\infty}$ is finite. Hence, the choice of exponential dependence of the second term, appropriate for the topological phase, is sufficient for our purposes (compare with Ref.~\cite{torlai18}). The~resulting $\delta_\infty$ is plotted as a function of disorder in Figure~\ref{fig:spectrum}{b}, and shows nicely the topological-to-trivial transition. The error bars for $\delta_\infty$ come from the error in determining the median $\delta_A$, which in turn was found using the bootstrap method.

We can now determine the phase transition point as the largest $W$ value where $\delta_\infty$ most probably vanishes. By this, we mean the value of $W$ at which the lower end of the $2\sigma$ confidence interval for $\delta_\infty$ approaches zero. This is determined by linear interpolation between the two probed values of $W$ between which this lower end of the confidence interval changes from positive to negative, as indicated in the inset to Figure~\ref{fig:spectrum}{b}. We can also define the largest $W$ which is certainly in the topological phase ($\delta_\infty=0$) as the one for which the error in $\delta_\infty$ is one order of magnitude larger than the estimated value of $\delta_\infty$. The distance between this point and the transition point will serve to set the error bar for the transition point.

\begin{figure}[t]
	\centering
	\includegraphics[width=\textwidth,clip]{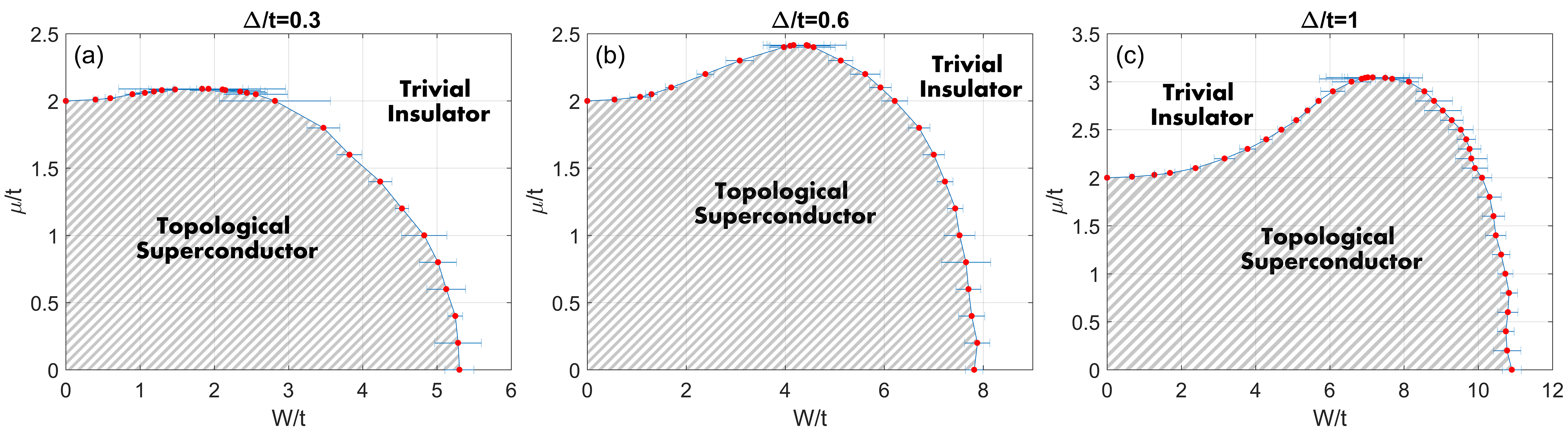}
	\caption{Phase diagram as a function of the disorder strength $W/t$ and the chemical potential $\mu/t$ for three values of $\Delta$: (\textbf{a}) $\Delta/t=0.3$, (\textbf{b}) $\Delta/t=0.6$, (\textbf{c}) $\Delta/t=1$. The hatched region is the topological superconductor phase, the other region is topologically trivial.}
	\label{fig:phase_diagram}
\end{figure}

Using this procedure, we determined the phase diagram of our system in the disorder strength-chemical potential plane for several values of $\Delta/t$. The results are displayed in Figure~\ref{fig:phase_diagram}. At the transition line, the entanglement entropy should display the critical logarithmic behavior, Equation~(\ref{eqn:s_cft}) with $c_\mathrm{eff} = \ln(2)/2$, except at the clean point ($W=0$ and $\mu/t=2$), where $c=1/2$. This is exactly the behavior found earlier in Figure~\ref{fig:entropy_mu} for several values of $\mu$ at $\Delta/t=0.6$. Let us note that the phase diagram of the system for $\Delta=t$ has been previously determined in reference~\cite{gregs16}, by looking for the presence of long range order in the $S^z$ correlation function in the spin version of the model, Equation~(\ref{eqn:ising}), and the results agree with ours. However, the possibility to identify the topological phase by appealing to some spin correlation function is very model-specific, while our method applies generally. Let us also note that reference~\cite{brzezicki17} studied the phase diagram of a tight-binding chain with a large unit cell in which pairing centers are distributed in an ordered or a disordered fashion. This allowed us to characterize the system using the Kitaev's topological index. However, this cannot be easily generalized to either the current case of a fully-disordered chain, or to the interacting regime. In contrast, our entanglement-based approach applies generically.

The most striking feature of the phase diagram is the existence of a topological region at $\mu/t>2$ for intermediate values of $W$, as anticipated in the previous section from the behavior of the entanglement entropy. In this region, the clean system is non-topological, and the topological phase is brought about by the disorder. Indeed, in the regime of large chemical potential, the clean system is almost completely filled with fermions, hence trivial. With disorder, regions of positive potential fluctuations can emerge, where a high density of holes may exist. These can create a network throughout the system, allowing for a topological phase with Majorana zero modes at its ends, as evidenced by our results.

\section{Conclusions} \label{sec:conclusions}
To conclude, in this work we have examined the interplay of a topological phase and disorder in the archetypical example of the random Kitaev chain, from the point of view of entanglement. We have found the length dependence of the entanglement entropy to change non-monotonically with disorder, indicating strong-disorder quantum criticality of the random transverse field Ising type, which persists even for $\Delta \ne t$, where the corresponding spin model is more complicated. We then developed a method to accurately identify the transition by looking at the entanglement spectrum and extracting from it the transition point and its uncertainty. Using this method, we verified the existence of a region in phase space where the topological phase is enabled by disorder.

In the future, it would be interesting to see how our results extend to interacting systems~\cite{crepin14,berkovits15,leiman15,gregs16,berkovits18,friedman18,berkovits18a} (where the density matrix renormalization group~\cite{white92,schollwoeck11} and related methods give direct access to the many-body density matrix and its spectrum), as well as to systems in higher dimensions~\cite{goldstein03,lian18}.


\vspace{6pt} 



\authorcontributions{Conceptualization, M.G., methodology, M.G. and L.L.; calculations, L.L.; analysis and writing, L.L. and M.G.}

\funding{This research was supported by the Israel Science Foundation (Grant No.~227/15), the German Israeli Foundation (Grant No.~I-1259-303.10), the US-Israel Binational Science Foundation (Grant No.~2016224), and the Israel Ministry of Science and Technology (Contract No.~3-12419).
}

\acknowledgments{We would like to thank R.~Berkovits, G.~Cohen, and E.~Sela for useful discussions.}

\conflictsofinterest{The authors declare no conflict of interest.} 

\reftitle{References}




\begin{thebibliography}{-------}
\providecommand{\natexlab}[1]{#1}

\bibitem[Amico \em{et~al.}(2008)Amico, Fazio, Osterloh, and Vedral]{amico08}
Amico, L.; Fazio, R.; Osterloh, A.; Vedral, V.
\newblock Entanglement in many-body systems.
\newblock {\em Rev. Mod. Phys.} {\bf 2008}, {\em 80},~517--576.
\newblock

\bibitem[Laflorencie(2016)]{laflorencie2016}
Laflorencie, N.
\newblock Quantum entanglement in condensed matter systems.
\newblock {\em Phys. Rep.} {\bf 2016}, {\em 646},~1--59.
\newblock

\bibitem[Kitaev(2001)]{kitaev01}
Kitaev, A.Y.
\newblock Unpaired Majorana fermions in quantum wires.
\newblock {\em Physics-Uspekhi} {\bf 2001}, {\em 44},~131.
\newblock

\bibitem[Alicea(2012)]{alicea12}
Alicea, J.
\newblock New directions in the pursuit of Majorana fermions in solid state
  systems.
\newblock {\em Rep. Prog. Phys.} {\bf 2012}, {\em 75},~076501.
\newblock

\bibitem[Sato and Fujimoto(2016)]{sato16}
Sato, M.; Fujimoto, S.
\newblock Majorana Fermions and Topology in Superconductors.
\newblock {\em J. Phys. Soc. Jpn.} {\bf 2016}, {\em
  85},~072001.
\newblock

\bibitem[Aguado(2017)]{aguado17}
Aguado, R.
\newblock Majorana quasiparticles in condensed matter.
\newblock {\em J. Phys. Soc.  Jpn.} {\bf 2017}, {\em
  40},~523--593.
\newblock

\bibitem[Alicea and Fendley(2016)]{alicea16}
Alicea, J.; Fendley, P.
\newblock Topological Phases with Parafermions: Theory and Blueprints.
\newblock {\em Annu. Rev. Condens. Matter Phys.} {\bf 2016}, {\em
  7},~119--139.
\newblock

\bibitem[Sarma \em{et~al.}(2015)Sarma, Freedman, and Nayak]{sarma15}
Sarma, S.D.; Freedman, M.; Nayak, C.
\newblock Majorana zero modes and topological quantum computation.
\newblock {\em Npj Quantum Inf.} {\bf 2015}, {\em 1},~15001.
\newblock

\bibitem[Lutchyn \em{et~al.}(2010)Lutchyn, Sau, and Das~Sarma]{lutchyn10}
Lutchyn, R.M.; Sau, J.D.; Das~Sarma, S.
\newblock Majorana Fermions and a Topological Phase Transition in
  Semiconductor-Superconductor Heterostructures.
\newblock {\em Phys. Rev. Lett.} {\bf 2010}, {\em 105},~077001.
\newblock

\bibitem[Oreg \em{et~al.}(2010)Oreg, Refael, and von Oppen]{oreg10}
Oreg, Y.; Refael, G.; von Oppen, F.
\newblock Helical Liquids and Majorana Bound States in Quantum Wires.
\newblock {\em Phys.~Rev. Lett.} {\bf 2010}, {\em 105},~177002.
\newblock

\bibitem[Mourik \em{et~al.}(2012)Mourik, Zuo, Frolov, Plissard, Bakkers, and
  Kouwenhoven]{mourik12}
Mourik, V.; Zuo, K.; Frolov, S.M.; Plissard, S.R.; Bakkers, E.P.A.M.;
  Kouwenhoven, L.P.
\newblock Signatures of Majorana Fermions in Hybrid
  Superconductor-Semiconductor Nanowire Devices.
\newblock {\em Science} {\bf 2012}, {\em 336},~1003--1007.
\newblock

\bibitem[Deng \em{et~al.}(2012)Deng, Yu, Huang, Larsson, Caroff, and
  Xu]{deng12}
Deng, M.T.; Yu, C.L.; Huang, G.Y.; Larsson, M.; Caroff, P.; Xu, H.Q.
\newblock Anomalous Zero-Bias Conductance Peak in a Nb–InSb Nanowire–Nb
  Hybrid Device.
\newblock {\em Nano Lett.} {\bf 2012}, {\em 12},~6414--6419.
\newblock

\bibitem[Das \em{et~al.}(2012)Das, Ronen, Most, Oreg, Heiblum, and
  Shtrikman]{das12}
Das, A.; Ronen, Y.; Most, Y.; Oreg, Y.; Heiblum, M.; Shtrikman, H.
\newblock Zero-bias peaks and splitting in an Al–InAs nanowire topological
  superconductor as a signature of Majorana fermions.
\newblock {\em Nat. Phys.} {\bf 2012}, {\em 8},~887--895.
\newblock

\bibitem[Finck \em{et~al.}(2013)Finck, Van~Harlingen, Mohseni, Jung, and
  Li]{finck13}
Finck, A.D.K.; Van~Harlingen, D.J.; Mohseni, P.K.; Jung, K.; Li, X.
\newblock Anomalous Modulation of a Zero-Bias Peak in a Hybrid
  Nanowire-Superconductor Device.
\newblock {\em Phys. Rev. Lett.} {\bf 2013}, {\em 110},~126406.
\newblock

\bibitem[Churchill \em{et~al.}(2013)Churchill, Fatemi, Grove-Rasmussen, Deng,
  Caroff, Xu, and Marcus]{churchill13}
Churchill, H.O.H.; Fatemi, V.; Grove-Rasmussen, K.; Deng, M.T.; Caroff, P.; Xu,
  H.Q.; Marcus, C.M.
\newblock Superconductor-nanowire devices from tunneling to the multichannel
  regime: Zero-bias oscillations and magnetoconductance crossover.
\newblock {\em Phys. Rev. B} {\bf 2013}, {\em 87},~241401.
\newblock

\bibitem[Li and Haldane(2008)]{li08}
Li, H.; Haldane, F.D.M.
\newblock Entanglement Spectrum as a Generalization of Entanglement Entropy:
  Identification of Topological Order in Non-Abelian Fractional Quantum Hall
  Effect States.
\newblock {\em Phys. Rev. Lett.} {\bf 2008}, {\em 101},~010504.
\newblock

\bibitem[Motrunich \em{et~al.}(2001)Motrunich, Damle, and Huse]{motrunich01}
Motrunich, O.; Damle, K.; Huse, D.A.
\newblock Griffiths effects and quantum critical points in dirty
  superconductors without spin-rotation invariance: One-dimensional examples.
\newblock {\em Phys. Rev. B} {\bf 2001}, {\em 63},~224204.
\newblock

\bibitem[Gruzberg \em{et~al.}(2005)Gruzberg, Read, and
  Vishveshwara]{gruzberg05}
Gruzberg, I.A.; Read, N.; Vishveshwara, S.
\newblock Localization in disordered superconducting wires with broken
  spin-rotation symmetry.
\newblock {\em Phys. Rev. B} {\bf 2005}, {\em 71},~245124.
\newblock

\bibitem[Akhmerov \em{et~al.}(2011)Akhmerov, Dahlhaus, Hassler, Wimmer, and
  Beenakker]{akhmerov11}
Akhmerov, A.R.; Dahlhaus, J.P.; Hassler, F.; Wimmer, M.; Beenakker, C.W.J.
\newblock Quantized Conductance at the Majorana Phase Transition in a
  Disordered Superconducting Wire.
\newblock {\em Phys. Rev. Lett.} {\bf 2011}, {\em 106},~057001.
\newblock

\bibitem[Fulga \em{et~al.}(2011)Fulga, Hassler, Akhmerov, and
  Beenakker]{fulga11}
Fulga, I.C.; Hassler, F.; Akhmerov, A.R.; Beenakker, C.W.J.
\newblock Scattering formula for the topological quantum number of a disordered
  multimode wire.
\newblock {\em Phys. Rev. B} {\bf 2011}, {\em 83},~155429.
\newblock

\bibitem[Potter and Lee(2011)]{potter11}
Potter, A.C.; Lee, P.A.
\newblock Engineering a $p+\mathit{ip}$ superconductor: Comparison of
  topological insulator and Rashba spin-orbit-coupled materials.
\newblock {\em Phys. Rev. B} {\bf 2011}, {\em 83},~184520.
\newblock

\bibitem[Stanescu \em{et~al.}(2011)Stanescu, Lutchyn, and
  Das~Sarma]{stanescu11}
Stanescu, T.D.; Lutchyn, R.M.; Das~Sarma, S.
\newblock Majorana fermions in semiconductor nanowires.
\newblock {\em Phys. Rev. B} {\bf 2011}, {\em 84},~144522.
\newblock

\bibitem[Brouwer \em{et~al.}(2011{\natexlab{a}})Brouwer, Duckheim, Romito, and
  von Oppen]{brouwer11a}
Brouwer, P.W.; Duckheim, M.; Romito, A.; von Oppen, F.
\newblock Topological superconducting phases in disordered quantum wires with
  strong spin-orbit coupling.
\newblock {\em Phys. Rev. B} {\bf 2011}, {\em 84},~144526.
\newblock

\bibitem[Brouwer \em{et~al.}(2011{\natexlab{b}})Brouwer, Duckheim, Romito, and
  von Oppen]{brouwer11b}
Brouwer, P.W.; Duckheim, M.; Romito, A.; von Oppen, F.
\newblock Probability Distribution of Majorana End-State Energies in Disordered
  Wires.
\newblock {\em Phys. Rev. Lett.} {\bf 2011}, {\em 107},~196804.
\newblock

\bibitem[Sau \em{et~al.}(2012)Sau, Tewari, and Das~Sarma]{sau12}
Sau, J.D.; Tewari, S.; Das~Sarma, S.
\newblock Experimental and materials considerations for the topological
  superconducting state in electron- and hole-doped semiconductors: Searching
  for non-Abelian Majorana modes in 1D nanowires and 2D heterostructures.
\newblock {\em Phys. Rev. B} {\bf 2012}, {\em 85},~064512.
\newblock

\bibitem[Lobos \em{et~al.}(2012)Lobos, Lutchyn, and Das~Sarma]{lobos12}
Lobos, A.M.; Lutchyn, R.M.; Das~Sarma, S.
\newblock Interplay of Disorder and Interaction in Majorana Quantum Wires.
\newblock {\em Phys. Rev. Lett.} {\bf 2012}, {\em 109},~146403.
\newblock

\bibitem[Pientka \em{et~al.}(2012)Pientka, Kells, Romito, Brouwer, and von
  Oppen]{pientka12}
Pientka, F.; Kells, G.; Romito, A.; Brouwer, P.W.; von Oppen, F.
\newblock Enhanced Zero-Bias Majorana Peak in the Differential Tunneling
  Conductance of Disordered Multisubband Quantum-Wire/Superconductor Junctions.
\newblock {\em Phys. Rev. Lett.} {\bf 2012}, {\em 109},~227006.
\newblock

\bibitem[Pientka \em{et~al.}(2013)Pientka, Romito, Duckheim, Oreg, and von
  Oppen]{pientka13}
Pientka, F.; Romito, A.; Duckheim, M.; Oreg, Y.; von Oppen, F.
\newblock Signatures of topological phase transitions in mesoscopic
  superconducting rings.
\newblock {\em New J. Phys.} {\bf 2013}, {\em 15},~025001.
\newblock

\bibitem[DeGottardi \em{et~al.}(2013)DeGottardi, Sen, and
  Vishveshwara]{degottardi13a}
DeGottardi, W.; Sen, D.; Vishveshwara, S.
\newblock Majorana Fermions in Superconducting 1D Systems Having Periodic,
  Quasiperiodic, and Disordered Potentials.
\newblock {\em Phys. Rev. Lett.} {\bf 2013}, {\em 110},~146404.
\newblock

\bibitem[Neven \em{et~al.}(2013)Neven, Bagrets, and Altland]{neven13}
Neven, P.; Bagrets, D.; Altland, A.
\newblock Quasiclassical theory of disordered multi-channel Majorana quantum
  wires.
\newblock {\em New J. Phys.} {\bf 2013}, {\em 15},~055019.
\newblock

\bibitem[Rieder \em{et~al.}(2013)Rieder, Brouwer, and Adagideli]{rieder13}
Rieder, M.T.; Brouwer, P.W.; Adagideli, I.
\newblock Reentrant topological phase transitions in a disordered spinless
  superconducting wire.
\newblock {\em Phys. Rev. B} {\bf 2013}, {\em 88},~060509.
\newblock

\bibitem[Sau and Das~Sarma(2013)]{sau13}
Sau, J.D.; Das~Sarma, S.
\newblock Density of states of disordered topological
  superconductor-semiconductor hybrid nanowires.
\newblock {\em Phys. Rev. B} {\bf 2013}, {\em 88},~064506.
\newblock

\bibitem[DeGottardi \em{et~al.}(2013)DeGottardi, Thakurathi, Vishveshwara, and
  Sen]{degottardi13b}
DeGottardi, W.; Thakurathi, M.; Vishveshwara, S.; Sen, D.
\newblock Majorana fermions in superconducting wires: Effects of long-range
  hopping, broken time-reversal symmetry, and potential landscapes.
\newblock {\em Phys. Rev. B} {\bf 2013}, {\em 88},~165111.
\newblock

\bibitem[Chevallier \em{et~al.}(2013)Chevallier, Simon, and Bena]{chevallier13}
Chevallier, D.; Simon, P.; Bena, C.
\newblock From Andreev bound states to Majorana fermions in topological wires
  on superconducting substrates: A story of mutation.
\newblock {\em Phys. Rev. B} {\bf 2013}, {\em 88},~165401.
\newblock

\bibitem[Jacquod and B\"uttiker(2013)]{jacquod13}
Jacquod, P.; B\"uttiker, M.
\newblock Signatures of Majorana fermions in hybrid normal-superconducting
  rings.
\newblock {\em Phys. Rev. B} {\bf 2013}, {\em 88},~241409.
\newblock

\bibitem[Adagideli \em{et~al.}(2014)Adagideli, Wimmer, and Teker]{adagideli14}
Adagideli, I.; Wimmer, M.; Teker, A.
\newblock Effects of electron scattering on the topological properties of
  nanowires: Majorana fermions from disorder and superlattices.
\newblock {\em Phys. Rev. B} {\bf 2014}, {\em 89},~144506.
\newblock

\bibitem[Hui \em{et~al.}(2014)Hui, Sau, and Das~Sarma]{hui14}
Hui, H.Y.; Sau, J.D.; Das~Sarma, S.
\newblock Generalized Eilenberger theory for Majorana zero-mode-carrying
  disordered $p$-wave superconductors.
\newblock {\em Phys. Rev. B} {\bf 2014}, {\em 90},~064516.
\newblock

\bibitem[Cr\'epin \em{et~al.}(2014)Cr\'epin, Zar\'and, and Simon]{crepin14}
Cr\'epin, F.M.C.; Zar\'and, G.; Simon, P.
\newblock Nonperturbative phase diagram of interacting disordered Majorana
  nanowires.
\newblock {\em Phys. Rev. B} {\bf 2014}, {\em 90},~121407.
\newblock

\bibitem[Gergs \em{et~al.}(2016)Gergs, Fritz, and Schuricht]{gregs16}
Gergs, N.M.; Fritz, L.; Schuricht, D.
\newblock Topological order in the Kitaev/Majorana chain in the presence of
  disorder and interactions.
\newblock {\em Phys. Rev. B} {\bf 2016}, {\em 93},~075129.
\newblock

\bibitem[Hegde and Vishveshwara(2016)]{hedge16}
Hegde, S.S.; Vishveshwara, S.
\newblock Majorana wave-function oscillations, fermion parity switches, and
  disorder in Kitaev chains.
\newblock {\em Phys. Rev. B} {\bf 2016}, {\em 94},~115166.
\newblock

\bibitem[Bagrets \em{et~al.}(2016)Bagrets, Altland, and Kamenev]{bagrets16}
Bagrets, D.; Altland, A.; Kamenev, A.
\newblock Sinai Diffusion at Quasi-1D Topological Phase Transitions.
\newblock {\em Phys. Rev. Lett.} {\bf 2016}, {\em 117},~196801.
\newblock

\bibitem[Grabsch and Texier(2016)]{grabsch16}
Grabsch, A.; Texier, C.
\newblock Topological phase transitions in the 1D multichannel Dirac equation
  with random mass and a random matrix model.
\newblock {\em Europhys. Lett.} {\bf 2016}, {\em 116},~17004.
\newblock

\bibitem[Pekerten \em{et~al.}(2017)Pekerten, Teker, Bozat, Wimmer, and
  Adagideli]{pekerten17}
Pekerten, B.; Teker, A.; Bozat, O.; Wimmer, M.; Adagideli, I.
\newblock Disorder-induced topological transitions in multichannel Majorana
  wires.
\newblock {\em Phys. Rev. B} {\bf 2017}, {\em 95},~064507.
\newblock

\bibitem[Brzezicki \em{et~al.}(2017)Brzezicki, Ole\ifmmode~\acute{s}\else
  \'{s}\fi{}, and Cuoco]{brzezicki17}
Brzezicki, W.; Ole\ifmmode~\acute{s}\else \'{s}\fi{}, A.M.; Cuoco, M.
\newblock Driving topological phases by spatially inhomogeneous pairing
  centers.
\newblock {\em Phys. Rev. B} {\bf 2017}, {\em 95},~140506.
\newblock

\bibitem[McGinley \em{et~al.}(2017)McGinley, Knolle, and
  Nunnenkamp]{mcginley17}
McGinley, M.; Knolle, J.; Nunnenkamp, A.
\newblock Robustness of Majorana edge modes and topological order: Exact
  results for the symmetric interacting Kitaev chain with disorder.
\newblock {\em Phys. Rev. B} {\bf 2017}, {\em 96},~241113.
\newblock

\bibitem[Lieu \em{et~al.}(2018)Lieu, Lee, and Knolle]{lieu18}
Lieu, S.; Lee, D.K.K.; Knolle, J.
\newblock Disorder protected and induced local zero-modes in longer-range
  Kitaev chains.
\newblock {\em Phys. Rev. B} {\bf 2018}, {\em 98},~134507.
\newblock

\bibitem[Monthus(2018)]{monthus18}
Monthus, C.
\newblock Topological phase transitions in random Kitaev $\alpha$-chains.
\newblock {\em J. Phys. A Math.  Theor.} {\bf 2018},
  {\em 51},~465301.
\newblock

\bibitem[Wang and Chakravarty(2018)]{wang18}
Wang, J.; Chakravarty, S.
\newblock Binary disorder in quantum Ising chains and induced Majorana zero
  modes.
\newblock {\em arXiv} {\bf 2018}, arXiv:1808.04481.

\bibitem[Mishra \em{et~al.}(2018)Mishra, Jafari, and Akbari]{mishra18}
Mishra, U.; Jafari, R.; Akbari, A.
\newblock Disordered Kitaev chain with long-range pairing: Loschimdt echo
  revivals and dynamical phase transitions.
\newblock {\em arXiv} {\bf 2018}, arXiv:1810.06236.

\bibitem[Peschel(2003)]{peschel03}
Peschel, I.
\newblock Calculation of reduced density matrices from correlation functions.
\newblock {\em J. Phys. A Math.  Gen.} {\bf 2003}, {\em
  36},~L205.
\newblock

\bibitem[Holzhey \em{et~al.}(1994)Holzhey, Larsen, and Wilczek]{holzhey94}
Holzhey, C.; Larsen, F.; Wilczek, F.
\newblock Geometric and renormalized entropy in conformal field theory.
\newblock {\em Nucl. Phys. B} {\bf 1994}, {\em 424},~443--467.
\newblock

\bibitem[Vidal \em{et~al.}(2003)Vidal, Latorre, Rico, and Kitaev]{vidal03}
Vidal, G.; Latorre, J.I.; Rico, E.; Kitaev, A.
\newblock Entanglement in Quantum Critical Phenomena.
\newblock {\em Phys. Rev. Lett.} {\bf 2003}, {\em 90},~227902.
\newblock

\bibitem[Calabrese and Cardy(2004)]{calabrese04}
Calabrese, P.; Cardy, J.
\newblock Entanglement entropy and quantum field theory.
\newblock {\em J. Stat. Mech. Theory  Exp.} {\bf
  2004}, {\em 2004},~P06002.
\newblock

\bibitem[Calabrese and Cardy(2009)]{calabrese09}
Calabrese, P.; Cardy, J.
\newblock Entanglement entropy and conformal field theory.
\newblock {\em J. Phys. A Math.  Theor.} {\bf 2009},
  {\em 42},~504005.
\newblock

\bibitem[Ma \em{et~al.}(1979)Ma, Dasgupta, and Hu]{ma79}
Ma, S.k.; Dasgupta, C.; Hu, C.k.
\newblock Random Antiferromagnetic Chain.
\newblock {\em Phys. Rev. Lett.} {\bf 1979}, {\em 43},~1434--1437.
\newblock

\bibitem[Dasgupta and Ma(1980)]{dasgupta80}
Dasgupta, C.; Ma, S.K.
\newblock Low-temperature properties of the random Heisenberg antiferromagnetic
  chain.
\newblock {\em Phys. Rev. B} {\bf 1980}, {\em 22},~1305--1319.
\newblock

\bibitem[Fisher(1994)]{fisher94}
Fisher, D.S.
\newblock Random antiferromagnetic quantum spin chains.
\newblock {\em Phys. Rev. B} {\bf 1994}, {\em 50},~3799--3821.
\newblock

\bibitem[Fisher(1995)]{fisher95}
Fisher, D.S.
\newblock Critical behavior of random transverse-field Ising spin chains.
\newblock {\em Phys. Rev. B} {\bf 1995}, {\em 51},~6411--6461.
\newblock

\bibitem[Refael and Moore(2004)]{refael04}
Refael, G.; Moore, J.E.
\newblock Entanglement Entropy of Random Quantum Critical Points in One
  Dimension.
\newblock {\em Phys. Rev. Lett.} {\bf 2004}, {\em 93},~260602.
\newblock

\bibitem[Refael and Moore(2009)]{refael09}
Refael, G.; Moore, J.E.
\newblock Criticality and entanglement in random quantum systems.
\newblock {\em J.  Phys. A Math.  Theor.} {\bf 2009},
  {\em 42},~504010.
\newblock

\bibitem[Torlai \em{et~al.}(2018)Torlai, McAlpine, and De~Chiara]{torlai18}
Torlai, G.; McAlpine, K.D.; De~Chiara, G.
\newblock Schmidt gap in random spin chains.
\newblock {\em Phys. Rev. B} {\bf 2018}, {\em 98},~085153.
\newblock

\bibitem[Berkovits(2015)]{berkovits15}
Berkovits, R.
\newblock Entanglement Properties and Quantum Phases for a Fermionic Disordered
  One-Dimensional Wire with Attractive Interactions.
\newblock {\em Phys. Rev. Lett.} {\bf 2015}, {\em 115},~206401.
\newblock

\bibitem[Leiman \em{et~al.}(2015)Leiman, Eisenbach, and Berkovits]{leiman15}
Leiman, S.; Eisenbach, A.; Berkovits, R.
\newblock Correspondence between many-particle excitations and the entanglement
  spectrum of disordered ballistic one-dimensional systems.
\newblock {\em Europhys. Lett.} {\bf 2015}, {\em 112},~46003.
\newblock

\bibitem[Berkovits(2018)]{berkovits18}
Berkovits, R.
\newblock Low eigenvalues of the entanglement Hamiltonian, localization length,
  and rare regions in one-dimensional disordered interacting systems.
\newblock {\em Phys. Rev. B} {\bf 2018}, {\em 97},~115408.
\newblock

\bibitem[Friedman and Berkovits(2018)]{friedman18}
Friedman, B.; Berkovits, R.
\newblock Entanglement entropy distribution in the strongly disordered
  one-dimensional Anderson model.
\newblock {\em arXiv} {\bf 2018}, arXiv:0812.0006.

\bibitem[Berkovits(2018)]{berkovits18a}
Berkovits, R.
\newblock Extracting many-particle entanglement entropy from observables using
  supervised machine learning.
\newblock {\em arXiv} {\bf 2018}, arXiv:1810.00346.

\bibitem[White(1992)]{white92}
White, S.R.
\newblock Density matrix formulation for quantum renormalization groups.
\newblock {\em Phys. Rev. Lett.} {\bf 1992}, {\em 69},~2863--2866.
\newblock

\bibitem[Schollwöck(2011)]{schollwoeck11}
Schollwöck, U.
\newblock The density-matrix renormalization group in the age of matrix product
  states.
\newblock {\em Ann. Phys.} {\bf 2011}, {\em 326},~96--192.

\bibitem[Goldstein and Berkovits(2003)]{goldstein03}
Goldstein, M.; Berkovits, R.
\newblock On-site interaction effects on localization: Dominance of
  nonuniversal contributions.
\newblock {\em Phys. Rev. B} {\bf 2003}, {\em 68},~245116.
\newblock

\bibitem[Lian \em{et~al.}(2018)Lian, Wang, Sun, Vaezi, and Zhang]{lian18}
Lian, B.; Wang, J.; Sun, X.Q.; Vaezi, A.; Zhang, S.C.
\newblock Quantum phase transition of chiral Majorana fermions in the presence
  of disorder.
\newblock {\em Phys. Rev. B} {\bf 2018}, {\em 97},~125408.
\newblock

\end{thebibliography}
\end{document}